\documentclass[aip,reprint,amsmath,amssymb]{revtex4-1}

\usepackage{graphicx}
\usepackage{dcolumn}
\usepackage{bm}
\usepackage{color}

\begin{document}

\title{Observation of fluctuation-induced tunneling conduction in micrometer-sized tunnel junctions}

\author{Yu-Ren Lai}
\affiliation{Institute of Physics, National Chiao Tung University, Hsinchu 30010, Taiwan}

\author{Kai-Fu Yu}
\affiliation{Institute of Physics, National Chiao Tung University, Hsinchu 30010, Taiwan}

\author{Yong-Han Lin}
\email{yonghanlin@gmail.com}
\affiliation{Institute of Physics, National Chiao Tung University, Hsinchu 30010, Taiwan}

\author{Jong-Ching Wu}
\affiliation{Department of Physics, National Changhua University of Education, Changhua 500, Taiwan}

\author{Juhn-Jong Lin}
\email{jjlin@mail.nctu.edu.tw}
\affiliation{Institute of Physics, National Chiao Tung University, Hsinchu 30010, Taiwan}
\affiliation{Department of Electrophysics, National Chiao Tung University, Hsinchu 30010, Taiwan}

\date{\today}

\begin{abstract}

Micrometer-sized Al/AlO$_{x}$/Y tunnel junctions were fabricated by the electron-beam lithography technique. The thin ($\approx$ 1.5--2 nm thickness) insulating AlO$_{x}$ layer was grown on top of the Al base electrode by O$_{2}$ glow discharge. The zero-bias conductances $G(T)$ and the current-voltage characteristics of the junctions were measured in a wide temperature range 1.5--300 K\@. In addition to the direct tunneling conduction mechanism observed in low-$G$ junctions, high-$G$ junctions reveal a distinct charge transport process which manifests the thermally fluctuation-induced tunneling conduction (FITC) through short nanoconstrictions. We ascribe the experimental realization of the FITC mechanism to originating from the formations of ``hot spots" (incomplete pinholes) in the AlO$_{x}$ layer owing to large junction-barrier interfacial roughness.

\end{abstract}


%\keywords{fluctuation-induced tunneling}

\maketitle

\section{Introduction}

\begin{figure}[bt]
\includegraphics[scale=0.15]{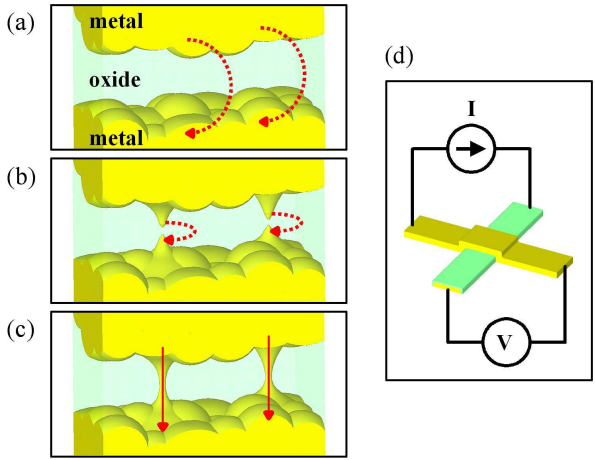}
\caption{\label{fig_1}%
(a)--(c) Schematic diagrams depicting three types of insulating barrier: (a) a comparatively uniform oxide barrier, (b) an oxide barrier with two hot spots (close approaches) but no pinholes, and (c) an oxide barrier with two pinholes. The red dashed or solid curves in each case schematically depicts the electron transport process. (d) Schematic diagram showing the four-probe measurement method on a tunnel junction.}
\end{figure}

Electron tunneling through a thin insulating layer (a potential barrier) in a metal-insulator-metal (MIM) multilayered structure is one of the most fundamental research topics in condensed matter physics. It also lies at the heart of numerous solid-state devices, such as tunnel diodes, \cite{Sze} Coulomb blockade thermometers, \cite{PekolaAPL00} Josephson junctions, \cite{Rippard02} and memory elements based on magnetic tunnel junctions. \cite{MooderaRPP11} The functionality of a tunnel device relies heavily on the material properties of the intermediate thin insulating layer. Usually, a weak insulating-like temperature dependence of the zero-bias junction conductance, $G(T) = [ \partial I(V,T)/\partial V ]_{V\,\rightarrow\,0}$, as described by the Simmons model \cite{SimmonsJAP63} is used to ascertain the quality and reliability of the insulating layer. \cite{SchullerAPL00} In the absence of any pinholes in the insulating layer [see the schematic diagram depicted in Fig.~\ref{fig_1}(a)], the Simmons model predicts a quadratic temperature dependent $G(T) \propto T^{2}$ law from low temperatures up to room temperature.

In practice, a precise control of the material properties of the thin insulating layer during the junction fabrication process is extremely difficult. Often, the junction yield varies sensitively with the oxide layer thickness. \cite{BarnerPRB89} Furthermore, the thickness distribution of the oxide layer may be substantial even for those junctions grown under similar conditions. \cite{PekolaJPCM03} As such, the $G(T)$ behavior might differ substantially from the Simmons law in certain cases. \cite{GuptaAPL97,GuptaJAP11} The key reason to explain these large variations in the MIM quality can be ascribed to the common formation of junction-barrier interfacial roughness. \cite{SchullerPRB06,SchullerAPL07} Because the electron transmission probability increases exponentially with decreasing barrier width, the presence of any notable interfacial roughness thus can significantly affect the overall junction transport properties.

In an MIM junction with marked interfacial roughness as schematically depicted in Fig.~\ref{fig_1}(b), the sharp points of the closest approaches (called ``hot spots" hereafter \cite{Freericks02}) between the two conducting regimes will no doubt play a predominant role in the electron tunneling. [We note that a hot spot is a precursor of a fully connected ``pinhole," see Fig.~\ref{fig_1}(c).] The current flowing through the hot spots between two large metal grains is theoretically found to depend sensitively not only on the applied bias voltage $V_{\rm a}$ but also on the thermal voltage $V_{\rm T}$ across the hot spots, where $V_{\rm T} = \pm \sqrt{k_{\rm B}T/C}$, with $k_{\rm B}$ being the Boltzmann constant, and $C$ being the capacitance of the hot spots. \cite{ShengPRL78,ShengPRB80} (The plus/minus sign in $V_{\rm T}$ stresses the fact that $V_{\rm T}$ can be in the same/opposite direction to $V_{\rm a}$.) Owing to the smallness of the hot spot, the magnitude of $C$ is tiny and thus $V_{\rm T}$ is notable. The resulting effect of the total voltage $V_{\rm total} = V_{\rm a} + V_{\rm T}$ gives rise to the so-called thermally fluctuation-induced tunneling conduction (FITC) process through the junction. The FITC model predicts unique functional forms for the zero-bias $G$ versus $T$ dependence [Eq.~(\ref{FITC_RT})] as well as finite-bias current-voltage ($I$-$V$) characteristics [Eq.~(\ref{FITC_IV})] over a wide range of $T$ from liquid-helium temperatures to room temperature. While this phenomenological FITC model has reasonably well described the $G(T)$ behavior and $I$-$V$ curves in a good number of experiments, such as nanowire contacts, \cite{LinNT08,LinJAP11} the extracted barrier height is often too small (e.g., $\lesssim$ few meV) and barrier width too large (e.g., $\gtrsim$ few tens nm). In order to resolve this puzzle, Xie and Sheng have recently retreated this problem by microscopically calculating the electronic wave transmission through atomic-scale fine structures (short ``nanoconstrictions") connecting two large metallic grains. \cite{XiePRB09} In particular, the relevant conduction channels they theoretically considered are those with a transverse dimension slightly smaller than the half of the Fermi wavelength of the tunneling electron. By employing the Landauer formula, they obtained the $G(T)$ attributes and $I$-$V$ characteristics very similar to the original phenomenological FITC predictions. In other words, short nanoconstrictions with a transverse dimension smaller than the cutoff wavelength (of a waveguide) can host thermally fluctuation induced tunneling conduction process at finite temperatures. \cite{XiePRB09} The extracted potential barrier height and width according to this microscopic model are predicted to be much more realistic.

On the experimental side, it is surprising that the possible occurrence of the FITC-like mechanism in MIM multilayered structures has never been reported, even though the formation of hot spots is very likely in certain junctions. In this paper, we present our observations of the FITC process governed $G(T)$ features and $I$-$V$ characteristics in micrometer-sized Al/AlO$_{x}$/Y tunnel junctions in a wide temperature range 1.5--300 K\@. The extracted parameters are examined in terms of the presence of hot spots in the AlO$_{x}$ layer, which mimics the short nanoconstrictions in the microscopic FITC model. Our results are presented below.

\section{Experimental method}

The fabrication method of our planar tunnel junctions was described previously. \cite{Sun-prb10} The designed geometrical junction area $A_{\rm d}$ ($\approx$ 20$\times$20 $\mu$m$^2$) was defined by the electron-beam lithography technique. Both the bottom Al ($\approx$ 30 nm thickness) and the top Y ($\approx$ 100 nm thickness) electrodes were deposited by thermal evaporation. Prior to the deposition of the top Y electrode, an insulating AlO$_{x}$ layer ($\approx$ 1.5--2 nm thickness on average) was grown on the top surface of the Al base electrode by the O$_{2}$ glow discharge. \cite{O2Plasma} Low angle x-ray diffraction studies revealed an amorphous structure of the oxide layer. The temperature dependent zero-bias resistance $R(T) = 1/G(T)$ and the $I$-$V$ curves of the junctions were measured by the four-probe method [Fig.~\ref{fig_1}(d)], using a Keithley K220 current source and a K182 voltmeter. A standard $^{4}$He cryostat equipped with a calibrated Si diode thermometer was utilized for $R(T)$ and $I$-$V$ curve measurements between 1.5 and 300 K\@. The details of our sample configuration and measurement method had previously been discussed in Ref. \onlinecite{Sun-prb10}.

We first discern the three possible types of junction-barrier interfacial roughness that can emerge in a fabricated MIM tunnel junction. Figure~\ref{fig_1}(a) depicts the normal case where the interfacial roughness is moderate. Electron tunneling occurs more or less evenly over the entire geometrical area $A_{\rm d}$, and the $T$ dependence of the junction conductance  $G$ obeys the well-established Simmons law. \cite{SimmonsJAP63} In the opposite case [Fig.~\ref{fig_1}(c)], pinholes may exist in the barrier, shorting the two electrodes, and cause overall metallic features of $G(T)$. What is more interesting and will be the focus of this paper is the intermediate case [Fig.~\ref{fig_1}(b)], where the interfacial roughness is notable but pinholes are just about to fully develop. Electron tunneling then occurs predominantly at the sharp points of the closest approaches of the top and bottom electrodes. Experimentally, the locations of such hot spots (incomplete pinholes) may be identified by mapping the tunnel current over the lateral area $A_{\rm d}$, e.g., from conducting atomic force microscopy study at room temperature. \cite{DaCostaJAP98,AndoJJAP99} Alternatively, the occurrence of such hot spots can substantiate the FITC process in a wide $T$ range from liquid-helium temperatures to room temperature, \cite{ShengPRL78,ShengPRB80} as we demonstrate below in micrometer-sized Al/AlO$_{x}$/Y tunnel junctions.

\section{Results and discussion}

According to the FITC model, \cite{ShengPRL78,ShengPRB80} the small effective junction area $A$ [i.e., the size of a hot spot depicted in Fig.~\ref{fig_1}(b), where $A \ll A_{\rm d}$] can lead to large random thermal voltages $V_{\rm T} = \pm \sqrt{k_{\rm B}T/C}$ that fluctuate across the narrow gap, which would in turn effectively lower and narrow the shape of the potential barrier. As a consequence, electron tunneling would be significantly influenced, causing $G(T)$ to increase exponentially with increasing $T$. (In the opposite limit of $T \rightarrow 0$, $G$ becomes constant, i.e., recovering the elastic tunneling behavior, due to the gradually vanishing $V_{\rm T}$ with $T$.) Previously, the FITC model has been applied to explain the $G(T)$ behavior observed in a good number of conductor-dielectric composite systems (granular films) \cite{DresselhausPRB94,KimSM01,LinNT08,LinJAP11,LongPPS11} and nanowire contacts. \cite{LinNT08,LinJAP11} Surprisingly, although the interfacial roughness must be very common in artificially fabricated MIM tunnel junctions, \cite{DaCostaJAP98,AndoJJAP99,BuchananAPL02,SchullerPRB06,SchullerAPL07} the possible manifestation of the FITC process through hot spots in the oxide layer has never been reported in the literature.

\begin{figure}[bt]
\includegraphics[scale=0.18]{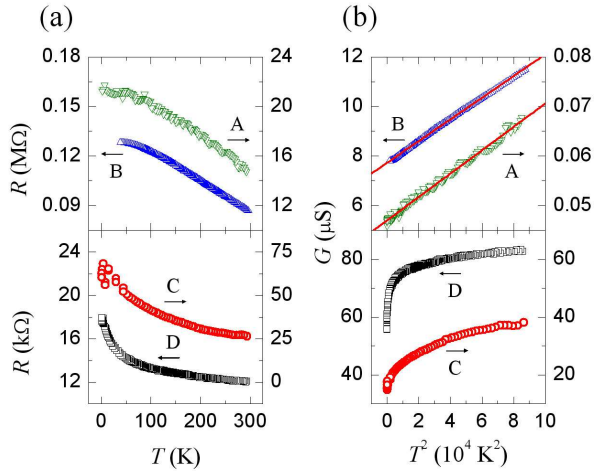}
\caption{\label{fig_2}%
(a) Zero-bias resistance $R$ as a function of $T$ and (b) the corresponding conductance $G$ as a function of $T^{2}$ for the junctions A to D, as indicated. The straight lines drawn through junctions A and B in the top panel of (b) are least-squares fits to Eq.~(\ref{Simmons}).}
\end{figure}

In this work, a dozen of Al/AlO$_{x}$/Y junctions have been fabricated and studied. In the following, we report the electrical-transport properties of four representative junctions. Two of them show the standard Simmons behavior, while the other two manifest the FITC mechanism. Figure~\ref{fig_2}(a) displays the $T$ dependence of $R$ for the four junctions A to D, as indicated. In the two higher resistance junctions A and B, $R$ increases only slightly with decreasing $T$ and it progressively saturates at $T$\,$\lesssim$\,100 K\@. In sharp contrast, in the two lower resistance junctions C and D, $R$ increases exponentially with decreasing $T$. The relative resistance ratios [$R(1.5~{\rm K})$$-$$R(300~{\rm K})$]/$R(300~{\rm K})$ in the junctions C and D are larger than those in the junctions A and B\@.

In the Simmons model, the zero-bias conductance, $G_{\rm S}(T)$, due to direct (elastic) tunneling of electrons through a potential barrier possesses a $T^2$ temperature dependence as given by  \cite{SimmonsJAP63}
\begin{equation}
G_{\rm S}(T)=G_{\rm S0}\left[ 1+\left( \frac{T}{T_{\rm S0}} \right)^{2} \right]~,%
\label{Simmons}%
\end{equation}
where $G_{\rm S0}$ and $T_{\rm S0}$ are temperature independent parameters. As shown in the top panel of Fig.~\ref{fig_2}(b), the $G \propto T^{2}$ law clearly holds for the junctions A and B in the wide temperature range 1.5--300 K\@. On the other hand, the bottom panel of Fig.~\ref{fig_2}(b) unambiguously  indicates that the conductances of the junctions C and D significantly deviate from the Simmons law.

In the phenomenological FITC theory, at small bias voltages (i.e., the ohmic $I$-$V$ regime), the $R(T)$ of a small junction is described by \cite{ShengPRL78,ShengPRB80}
\begin{equation}
R(T)=R_{\infty}\,{\rm exp}\left(\frac{T_{1}}{T_{0}+T}\right)~,%
\label{FITC_RT}%
\end{equation}
where the parameter $R_{\infty}$ depends only weakly on $T$, and the characteristic temperatures
\begin{equation}
T_{1} = \frac{8 \varepsilon_{\rm r} \varepsilon_{0} A \phi_{0}^{2}}{k_{\rm B} e^{2} w}%
\label{T1}
\end{equation}
and
\begin{equation}
T_{0} = \frac{16 \varepsilon_{\rm r} \varepsilon_{0} \hbar A \phi_{0}^{3/2}}{\pi (2m)^{1/2} k_{\rm B} e^{2} w^{2}}~,%
\label{T0}
\end{equation}
where $\phi_{0}$ is the barrier height, $w$ is the barrier width, $\varepsilon_{0}$ is the permittivity of vacuum, $\varepsilon_{\rm r}$ is the dielectric constant of the insulating barrier (for Al$_{2}$O$_{3}$, $\varepsilon_{\rm r} \approx$ 4.5--8.4, see Ref.~\onlinecite{CRC}), $2\pi\hbar$ is the Planck's constant, and $m$ is the electronic mass. The characteristic thermal energy $k_{\rm B}T_{1}$ can be regarded as a measure of the energy required for an electron to cross the barrier, and $T_{0}$ is the temperature well below which fluctuation effects become insignificant.

\begin{figure}[bt]
\includegraphics[scale=0.18]{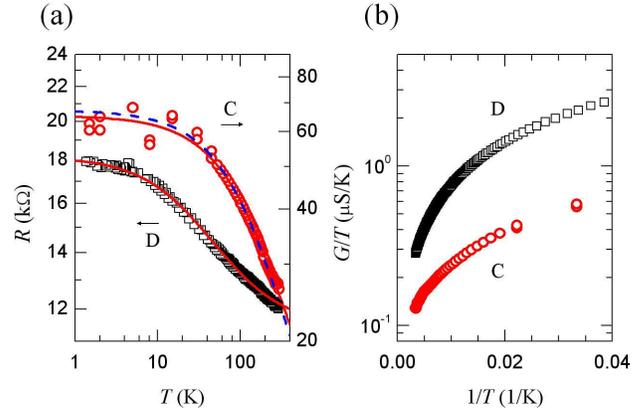}
\caption{\label{fig_3}%
(a) Log-log plot of $R$ versus $T$ for junctions C and D, as indicated. The solid curves are least-squares fits to Eq.~(\ref{FITC_RT}). The dashed curve through junction C is the theoretical prediction of Eq.~(\ref{FITC_RT}) but is plotted by directly substituting the $T_{1}$ and $T_{0}$ values extracted from $a(T)$ in Eq.~(\ref{FITC_IV}). (b) Log($G/T$) versus $1/T$ for junctions C and D between 25 and 300 K\@. Note that there exists no linear regime in any temperature regime.}
\end{figure}

Figure~\ref{fig_3}(a) shows a log-log plot of the measured $R$ as a function of $T$ for the junctions C and D\@. The symbols are the experimental data and the solid curves are the least-squares fits to Eq.~(\ref{FITC_RT})\@. The fitted $R_{\infty}$, $T_{1}$, and $T_{0}$ values are listed in Table~\ref{table_1}. The predictions of Eq.~(\ref{FITC_RT}) are seen to well describe the data. (The dashed curve drawn through the junction C will be discussed below.)

\begin{table}[bt]
\caption{\label{table_1}%
Parameters for junctions C and D\@. The values of $w$ and $\phi_{0}$ were extracted by using the geometrical area $A_{\rm d}$ ($\simeq$ $20 \times 20$ $\mu$m$^2$), and the dielectric constant $\varepsilon_{\rm r}$(Al$_{2}$O$_{3}$) $\approx$ 6.}
\begin{ruledtabular}
\begin{tabular}{ccccccc}
Junction  &Method  &$R_{\infty}$  &$T_{1}$  &$T_{0}$  &$w$  &$\phi_{0}$
\\
&  &(k$\Omega$)  &(K)  &(K)  &(nm)  &(meV)
\\
\hline
C  &$R(T)$                        &7.66  &815   &380   &37.7  &0.050\\
   &$I_{\rm h}(V_{\rm h})$        &6.46  &915   &391   &39.5  &0.054\\
D  &$R(T)$                        &11.4  &21.5  &46.5  &22.9  &0.006\\
\end{tabular}
\end{ruledtabular}
\end{table}

It might be conjectured that, if $\phi_{0}$ is small, the exponential increase of $R$ with decreasing $T$ in the junctions C and D could arise from the thermionic emission of electrons over the barrier. \cite{TsaiAPL06,Sze} If this were the case, one would expect the total conductance to be given by the sum of the Simmons conductance and the thermionic-emission conductance $G_{\rm t}(T)$, i.e., $G(T)$\,$=$\,$G_{\rm S}(T)$\,$+$\,$G_{\rm t}(T)$, where
$G_{\rm t}(T)=G_{\rm t0}\,T\,{\rm exp} (- T_{\rm t0} /T)$, with $G_{\rm t0}$ and $T_{\rm t0}$ being two temperature independent parameters. At low temperatures, $G_{\rm S}(T)$ should dominate over $G_{\rm t}(T)$\@. As $T$ increases, $G_{\rm t}(T)$ would become progressively important and might eventually dominate over $G_{\rm S}(T)$\@. Figure~\ref{fig_3}(b) shows a plot of log$(G/T)$ versus $1/T$ for the junctions C and D\@. Clearly, there does not exist any linear regime between 25 and 300 K\@. Therefore, the thermionic-emission conduction process can not account for our measured $G$ versus $T$ dependence.

We return to the FITC mechanism. As recently discussed in Ref.~\onlinecite{LinJAP11} by two of the authors, one may alternatively extract the values of $T_{1}$ and $T_{0}$ by measuring the non-ohmic $I$-$V$ characteristics of a given junction at high bias voltages $V_{\rm h}$. Under such conditions, the phenomenological FITC theory predicts a nonlinear $I$-$V$ curve given by: \cite{ShengPRL78,ShengPRB80}
\begin{equation}
I_{\rm h}(V_{\rm h},T)=I_{\rm hs}\,%
{\rm exp}\left[-a(T) \left(1-\frac{V_{\rm h}}{V_{\rm hc}}\right)^{2}\,\right]~,%
~~~|V_{\rm h}|<V_{\rm hc}%
\label{FITC_IV}%
\end{equation}
where the saturation current $I_{\rm hs}$ and the critical voltage $V_{\rm hc}$ are two parameters which depend only weakly on $T$. The parameter $a(T) = T_{1}/(T_{0}+T)$ describes the temperature effect on the $I$-$V$ curves. The value of $a(T)$ at each temperature can be extracted by fitting the measured $I_{\rm h}(V_{\rm h})$ curve to Eq.~(\ref{FITC_IV}). Then, the values of $T_{1}$ and $T_{0}$ can be inferred by fitting $a(T)$ to the expression $T_{1}/(T_{0}+T)$. Note that with the two sets of $T_{1}$ and $T_{0}$ values deduced from the two complementary manners [Eq.~(\ref{FITC_RT}) and Eq.~(\ref{FITC_IV})], one may perform a quantitative self-consistency check of the FITC theoretical predictions.

\begin{figure}[bt]
\includegraphics[scale=0.18]{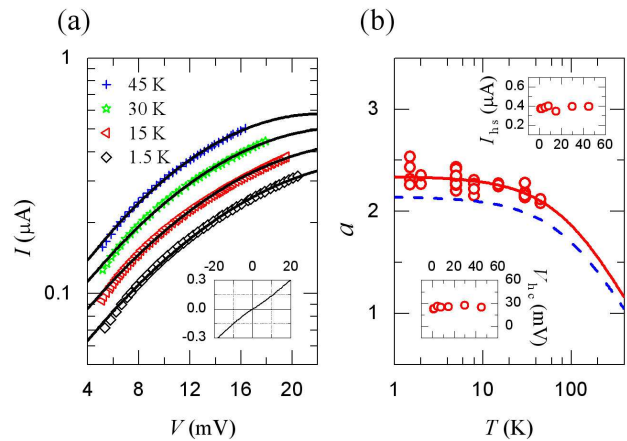}
\caption{\label{fig_4}%
Current-voltage characteristics of junction C: (a) nonlinear $I$-$V$ curves at four $T$ values. The solid curves are least-squares fits to Eq.~(\ref{FITC_IV})\@. For clarity, the data for 15, 30, and 45 K have been shifted up by multiplying factors of 1.2, 1.4, and 1.6, respectively. Inset: the $I$-$V$ curve at 1.5 K in double-linear scales. (b) The parameter $a$ in Eq.~(\ref{FITC_IV}) as a function of $T$. The solid curve is the least-squares fits to $T_{1}$/($T_{0}$\,+\,$T$). The dashed curve is the theoretical prediction of $T_{1}$/($T_{0}$\,+\,$T$) but is plotted by directly substituting the $T_{1}$ and $T_{0}$ values extracted from the $R(T)$ fit to Eq.~(\ref{FITC_RT})\@. Insets: (top) variation of $I_{\rm hs}$ and (bottom) variation of $V_{\rm hc}$ with $T$.}
\end{figure}

Figures~\ref{fig_4}(a) shows the measured nonlinear $I$-$V$ curves for the junction C at four $T$ values. The symbols are the experimental data and the solid curves are the least-squares fits to Eq.~(\ref{FITC_IV})\@. A good agreement between the theory and experiment is evident. Our extracted values of $a$ as a function of $T$ are plotted in Fig.~\ref{fig_4}(b), together with the least-squares fits (solid curves) to the expression $T_{1}/(T_{0}+T)$. The $T_{1}$ and $T_{0}$ values thus inferred are listed in Table~\ref{table_1}. The insets of Fig~\ref{fig_4}(b) plot the variations of our fitted $I_{\rm hs}$ and $V_{\rm hc}$ values with $T$. Note that they are essentially temperature independent, as predicted by the FITC theory.

We now substitute those $T_{1}$ and $T_{0}$ values extracted from the $R(T)$ fit [Eq.~(\ref{FITC_RT})] into the expression $T_{1}/(T_{0}+T)$ to compute $a(T)$ for the junction C\@. The calculated result (dashed curve) is plotted in Fig.~\ref{fig_4}(b), which is seen to describe the $a(T)$ behavior reasonably well. In turn, in Fig.~\ref{fig_3}(a) we plot the prediction of Eq.~(\ref{FITC_RT}) by directly substituting those $T_{1}$ and $T_{0}$ values inferred from the $a(T)$ fit through Eq.~(\ref{FITC_IV}) (while allowing $R_{\infty}$ as the sole adjustable parameter, whose value is listed in Table~\ref{table_1}). The result (dashed curve) is clearly in excellent agreement with the experimental data. Numerically, our $T_{1}$ and $T_{0}$ values extracted from the two methods differ by an amount of $\approx$\,10\%, which is very satisfactory and notably smaller than those ($\approx$\,30--40\%) reported in previous studies. \cite{KimSM01,LinJAP11} This close consistency provides a strong support for the pivotal relevance of the FITC mechanism occurring in our micrometer-sized tunnel junctions.

We would like to comment on our inferred values of the barrier parameters $\phi_{0}$ and $w$. Inspection of Table~\ref{table_1} indicates that the $\phi_{0}$ values are relatively small ($\lesssim$ 0.05 meV), while the $w$ values are relatively large ($\gtrsim$ 20 nm). These seemingly nonphysical values can be reconciled by taking into account the possible formation of hot spots as those depicted in Fig.~\ref{fig_1}(b). Since the hot-spot area $A$\,$\ll$\,$A_{\rm d}$, our using the geometrical area $A_{\rm d}$ to extract $\phi_{0}$ and $w$ from the inferred $T_{1}$ and $T_{0}$ values through Eqs.~(\ref{T1}) and (\ref{T0}) hence led to huge underestimate (overestimate) of $\phi_{0}$ ($w$). Although the actual area $A$ of a hot spot can not be precisely known for a given tunnel junction, it has previously been pointed out that the typical size of a hot spot can be about a few nanometers. \cite{DaCostaJAP98,AndoJJAP99} Then, taking $A$\,$\approx$ 2$\times$2 nm$^{2}$, one obtains values of $\phi_{0}$ $\approx$ 80 (10) meV and $w$ $\approx$ 0.9 (0.6) nm for the junction C (D)\@. These values are much acceptable.

In fact, the relevant hot-spot area could even be reduced to the sub-nanometer scale. In order to search for a microscopic insight and to obtain meaningful parameter values, Xie and Sheng \cite{XiePRB09} have recently considered the electron tunneling through finite segment(s) of nanoconstrictions whose transverse dimension is less than the half of the electronic Fermi wavelength $\lambda_{\rm F}$. In this new theoretical proposal, the insulating gap in the original FITC model \cite{ShengPRB80} is replaced by short and very narrow constrictions (which behave like waveguides with transverse dimensions slightly shorter than the cutoff wavelength $\lambda_{\rm F}/2$). By applying the Landauer formula and taking into account the effects of thermally induced voltage fluctuations across the nanoconstrictions, Xie and Sheng obtained a $G$ versus $T$ dependence very similar to the original FITC prediction [Eq.~(\ref{FITC_RT})] but with more realistic parameter values. Following this new theoretical idea and taking $A$\,$\approx$ 0.2$\times$0.2 nm$^{2}$ [note that $\lambda_{\rm F}$(Al) $\approx$ 0.36 nm], we obtain $\phi_{0}$ $\approx$ 0.5 eV and $w$ $\approx$ 0.4 nm for the junction C\@. These values are fairly realistic. Therefore, the manifestation of the FITC mechanism in micrometer-sized tunnel junctions implies the existence of such fine structures (hot spots or short nanoconstrictions) in the oxide layer. Electron tunneling through these fine structures governs the overall $G(T)$ behavior. [There could be a few such fine structures coexisting in a given MIM junction. However, the measured $G(T)$ might be predominated by only one of them, owing to the exponential dependence of the electron transmissivity on the fine-structure geometries.]

In a series of ballistic electron emission microscopy (BEEM) and scanning tunneling spectroscopy (STS) studies, Buhrman and coworkers \cite{Rippard02,Buhrman} have systematically shown that in an ultrathin ($\lesssim$ 1 nm) and not fully oxidized AlO$_x$ layer (which was not covered with a top electrode), there could possibly exist low-energy electron channels that could provide low-voltage ``leakage" through the barrier. In the present study, our AlO$_x$ layers are relatively thicker (1.5--2 nm) and fully oxidized by the O$_2$ glow discharge, as mentioned in Sec. II. Therefore, we think that the features of the FITC conduction process observed in our Al/AlO$_x$/Y junctions are not associated with the possible presence of low-energy electron channels that extended through the thin disordered oxide layer.

Theoretically, the crossover from a tunneling regime to an incoherent thermal-activation regime in a correlated barrier has been investigated by Freericks and coauthors. \cite{Freericks04} They have shown generically that the crossover takes place around a characteristic temperature which approximately equals a generalized Thouless energy for the barrier. Under certain parameter values, the predicted temperature behavior of the junction resistance (or, the specific resistance $RA_{\rm d}$) looks nominally similar to the temperature dependence shown in Fig. \ref{fig_3}(a). Unfortunately, the theory of Freericks does not provide a functional form for comparison with experiment. This issue deserves further studies.

\section{Conclusion}

We have observed the fluctuation-induced tunneling conduction mechanism in micrometer-sized Al/AlO$_{x}$/Y tunnel junctions in a wide temperature range 1.5--300 K\@. The manifestation of the FITC process at the micrometer scale reflects the existence of large junction-barrier interfacial roughness in the thin oxide layer, where electron tunneling occurs at the sharp points of the closest approaches of the top and bottom electrodes. The formation of a few hot spots, but not pinholes, predominantly governs the conductance versus temperature behavior as well as the current-voltage characteristics. Our results can have important bearing on the reliability and functionality of solid-state tunnel devices.

\begin{acknowledgments}

The authors are grateful to Ping Sheng for helpful discussion. This work was supported by the Taiwan National Science Council through Grant No. NSC 100-2120-M-009-008 and by the MOE ATU Program.

\end{acknowledgments}


\begin{thebibliography}{99}

\bibitem{Sze} S. M. Sze and K. K. Ng, {\it Physics of Semiconductor Devices}, 3rd ed. (Wiley Interscience, Hoboken, 2007).

\bibitem{PekolaAPL00} K. Gloos, R. S. Poikolainen, and J. P. Pekola, Appl. Phys. Lett. {\bf 77}, 2915 (2000).

\bibitem{Rippard02} W. H. Rippard, A. C. Perrella, F. J. Albert, and R. A. Buhrman, Phys. Rev. Lett. \textbf{88}, 046805 (2002).

\bibitem{MooderaRPP11} G. X. Miao, M. M{\"u}nzenberg, and J. S. Moodera, Rep. Prog. Phys. {\bf 74}, 036501 (2011).

\bibitem{SimmonsJAP63} J. G. Simmons, J. Appl. Phys. {\bf 35}, 2655 (1963).

\bibitem{SchullerAPL00} B. J. J{\"o}nsson-{\AA}kerman, R. Escudero, C. Leighton, S. Kim, I. K. Schuller, and D. A. Rabson, Appl. Phys. Lett. {\bf 77}, 1870 (2000).

\bibitem{BarnerPRB89} J. B. Barner and S. T. Ruggiero, Phys. Rev. B {\bf 39}, 2060 (1989).

\bibitem{PekolaJPCM03} K. Gloos, P. J. Koppinen, and J. P. Pekola, J. Phys.: Condens. Matter {\bf 15}, 1733 (2003).

\bibitem{GuptaAPL97} J. Z. Sun, L. Krusin-Elbaum, P. R. Duncombe, A. Gupta, and R. B. Laibowitz, Appl. Phys. Lett. {\bf 70}, 1769 (1997).

\bibitem{GuptaJAP11} M. Pathak, D. Mazumdar, V. Karthik, X. Zhang, K. B. Chetry, S. Keshavarz, P. LeClair, and A. Gupta, J. Appl. Phys. {\bf 110}, 053708 (2011).

%\bibitem{TuschePRL05} C. Tusche, H. L. Meyerheim, N. Jedrecy, G. Renaud, A. Ernst, J. Henk, P. Bruno, and J. Kirschner, Phys. Rev. Lett. {\bf 95}, 176101 (2005).

\bibitem{SchullerPRB06} C. W. Miller, Z. P. Li, I. K. Schuller, R. W. Dave, J. M. Slaughter, and J. {\AA}kerman, Phys. Rev. B {\bf 74}, 212404 (2006).

\bibitem{SchullerAPL07} C. W. Miller, Z. P. Li, J. {\AA}kerman, and I. K. Schuller, Appl. Phys. Lett. {\bf 90}, 043513 (2007).

\bibitem{Freericks02} A hot spot was called an intrinsic pinhole effect in  J. K. Freericks, B. K. Nikoli\'{c}, and P. Miller, Int. J. Mod. Phys. B \textbf{16}, 531 (2002).

\bibitem{ShengPRL78} P. Sheng, E. K. Sichel, and J. I. Gittleman, Phys. Rev. Lett. {\bf 40}, 1197 (1978).

\bibitem{ShengPRB80} P. Sheng, Phys. Rev. B {\bf 21}, 2180 (1980).

\bibitem{LinNT08} Y. H. Lin, S. P. Chiu, and J. J. Lin, Nanotechnology {\bf 19}, 365201 (2008).

\bibitem{LinJAP11} Y. H. Lin and J. J. Lin, J. Appl. Phys. {\bf 110}, 064318 (2011).

\bibitem{XiePRB09} H. Xie and P. Sheng, Phys. Rev. B {\bf 79}, 165419 (2009).

\bibitem{Sun-prb10} Y. C. Sun, S. S. Yeh, and J. J. Lin, Phys. Rev. B {\bf 82}, 054203 (2010).

%\bibitem{YehPRB09} S. S. Yeh and J. J. Lin, Phys. Rev. B {\bf 79}, 012411 (2009).

\bibitem{O2Plasma} J. L. Miles and P. H. Smith, J. Electrochem. Soc. {\bf 110}, 1240 (1963).

\bibitem{DaCostaJAP98} V. Da Costa, F. Bardou, C. B{\'e}al, Y. Henry, J. P. Bucher, and K. Ounadjela, J. Appl. Phys. {\bf 83}, 6703 (1998).

\bibitem{AndoJJAP99} Y. Ando, H. Kameda, H. Kubota, and T. Miyazaki, Jpn. J. Appl. Phys. {\bf 38}, L737 (1999).

\bibitem{DresselhausPRB94} Z. H. Wang, M. S. Dresselhaus, G. Dresselhaus, K. A. Wang, and P. C. Eklund, Phys. Rev. B {\bf 49}, 15890 (1994).

\bibitem{KimSM01} G. T. Kim, S. H. Jhang, J. G. Park, Y. W. Park, and S. Roth, Synth. Met. {\bf 117}, 123 (2001).

\bibitem{LongPPS11} Y. Z. Long, M. M. Li, C. Gu, M. Wan, J. L. Duvail, Z. Liu, and Z. Fan, Prog. Polym. Sci. {\bf 36}, 1415 (2011).

\bibitem{BuchananAPL02} J. D. R. Buchanan, T. P. A. Hase, B. K. Tanner, N. D. Hughes, and R. J. Hicken, Appl. Phys. Lett. {\bf 81}, 751 (2002).

\bibitem{CRC} {\it CRC Handbook of Tables for Applied Engineering Science}, 2nd ed., edited by R. E. Bolz and G. L. Tuve (CRC Press, Boca Raton, 1983).

\bibitem{TsaiAPL06} H. Im, Yu. A. Pashkin, T. Yamamoto, O. Astafiev, Y. Nakamura, and J. S. Tsai, Appl. Phys. Lett. {\bf 88}, 112113 (2006).

\bibitem{Buhrman} A. C. Perrella, W. H. Rippard, P. G. Mather, M. J. Plisch, and R. A. Buhrman, Phys. Rev. B \textbf{65}, 201403(R) (2002); E. Tan, P. G. Mather, A. C. Perrella, J. C. Read, and R. A. Buhrman, Phys. Rev. B \textbf{71}, 161401(R) (2005); P. G. Mather, A. C. Perrella, E. Tan, J. C. Read, and R. A. Buhrman, Appl. Phys. Lett. \textbf{86}, 242504 (2005).

\bibitem{Freericks04} J. K. Freericks, Appl. Phys. Lett. \textbf{84}, 1383 (2004); Phys. Rev. B \textbf{70}, 195342 (2004).

%\bibitem{BeasleyPRB95} Y. Xu, D. Ephron, and M. R. Beasley, Phys. Rev. B {\bf 52}, 2843 (1995).

\end{thebibliography}
\end{document}